\documentclass[twocolumn,aps,prl,showpacs]{revtex4-1}
\usepackage[T1]{fontenc}
\usepackage[latin9]{inputenc}
\usepackage{amsmath}
\usepackage{amssymb}
\usepackage{graphicx}

\makeatletter
\@ifundefined{textcolor}{}
{%
 \definecolor{BLACK}{gray}{0}
 \definecolor{WHITE}{gray}{1}
 \definecolor{RED}{rgb}{1,0,0}
 \definecolor{GREEN}{rgb}{0,1,0}
 \definecolor{BLUE}{rgb}{0,0,1}
 \definecolor{CYAN}{cmyk}{1,0,0,0}
 \definecolor{MAGENTA}{cmyk}{0,1,0,0}
 \definecolor{YELLOW}{cmyk}{0,0,1,0}
}

\usepackage{dcolumn}
\usepackage{bm}
\usepackage{float}

\makeatother

\begin{document}

\title{From Explosive to Infinite-Order Transitions on a Hyperbolic Network}

\author{Vijay Singh, C.~T. Brunson and Stefan Boettcher}

\affiliation{Department of Physics, Emory University, Atlanta, GA, 30322; USA}
\begin{abstract}
We analyze the phase transitions that emerge from the recursive design of certain hyperbolic networks 
that includes, for instance, a discontinuous ("explosive")  transition in ordinary percolation. To this end, we solve the $q$-state
Potts model in the analytic continuation for non-integer $q$ with the real-space renormalization group.
We find exact expressions for this one-parameter family of models that describe the dramatic 
transformation of the transition. In particular, this variation in $q$ shows that the discontinuous 
transition is generic  in the regime $q<2$ that includes percolation. A continuous ferromagnetic 
transition is recovered  in a singular manner \emph{only} for the Ising model, $q=2$. For $q>2$ the 
transition immediately transforms into an infinitely smooth order parameter 
of the Berezinskii-Kosterlitz-Thouless (BKT) type.
\end{abstract}

\maketitle
Real-world networks~\citep{Barthelemy03,Boccaletti06,Dorogovtsev08} exhibit dramatically distinct
phenomenology, with a profound imprint of their geometry on the dynamics, when compared with lattices or mean-field systems.
What we now call complex networks, aside from being random, possess
geometries dominated by small-world bonds and scale-free degree distributions~\citep{Watts98,Barabasi99}. These lead to novel, and often non-universal,
scaling behaviors unknown for lattices. For example, 
they  exhibit discontinuous transition even in ordinary
percolation~\citep{Boettcher11d} as exactly solvable realizations
of the so-called ``explosive'' percolation transitions~\citep{Achlioptas09,ISI:000286751500010,ISI:000291093600009,PhysRevE.84.020101,Riordan11,Bastas14,Araujo14}. Unlike the Achlioptas process, it is easy to prove that this discontinuity exists~\citep{Boettcher11d,Singh14}, however, its origin results from the network structure, not from a correlated process~\citep{Achlioptas09}.

Hyperbolic  networks combine a lattice geometry with a hierarchy of 
small-world connections. They were originally proposed as solvable models for complex 
networks~\citep{Hinczewski06,SWPRL,Boettcher09c,Nogawa09,PhysRevLett.108.255703}.
These recursively defined structures provide deeper insights into small-world effects compared to 
random networks that otherwise require approximate or numerical methods.  Work on 
percolation~\citep{Boettcher09c,Berker09,PhysRevE.82.011113,Boettcher11d,Hasegawa13c},
the Ising model~\citep{Bauer05,Hinczewski06,Boettcher10c,Baek11}, and the $q$-state Potts 
model~\citep{Nogawa12,PhysRevLett.108.255703} have shown that critical behavior, once thought to 
be exotic and model-specific~\citep{Dorogovtsev08}, can be universally categorized near the transition 
point~\citep{BoBr12,Nogawa12} for a large class of hyperbolic networks, such as  those discontinuous 
percolation transitions described in Refs.~\citep{Boettcher11d,Singh14}. They exemplify a non-trivial 
category of the explosive cluster-growth mechanisms leading to the discontinuous 
transition~\citep{Cho14}.  Yet, it has been noted that this discontinuity arises only as a non-generic
case of the general theory~\citep{BoBr12}, making its prevalence in percolation on hyperbolic networks 
somewhat surprising. 

Here, we illuminate the origin of the explosive percolation transition by investigating the $q$-state Potts 
model on a simple hyperbolic network, MK1~\citep{Boettcher11d}. The analytic continuation in
$q$~\citep{Wu82} reveals that the discontinuity that characterizes the explosive percolation transition 
related to the limit $q\to1$, rather than being a special case, persists for all $q<2$. Instead, the Ising 
ferromagnet ($q=2$) emerges as the singular case between all discontinuous ($q<2$) and all 
infinite-order transitions ($q>2$), with an infinitely smooth order parameter of the 
Berezinskii-Kosterlitz-Thouless (BKT) type that is consistent with the theory in Ref.~\citep{BoBr12}.
In particular, with real-space renormalization group (RG), we trace the existence of the
discontinuous transition to the non-generic behavior of the critical exponent $y_{h}$ that describes the 
response of the system to a conjugate external field~\citep{Boettcher11d,PhysRevLett.108.255703}. This exponent, that is temperature-dependent in these hyperbolic networks, happens to reach its maximum exactly at the critical temperature for $q<2$, resulting in merely quadratic corrections in its local expansion. We argue that this exceptional behavior for $q<2$ is intimately connected with the manner in which geometric, lattice-like features are coupled with the hyperbolic structure.

\begin{figure}
\includegraphics[bb=0bp 0bp 250bp 560bp,clip,angle=270,width=1\columnwidth]{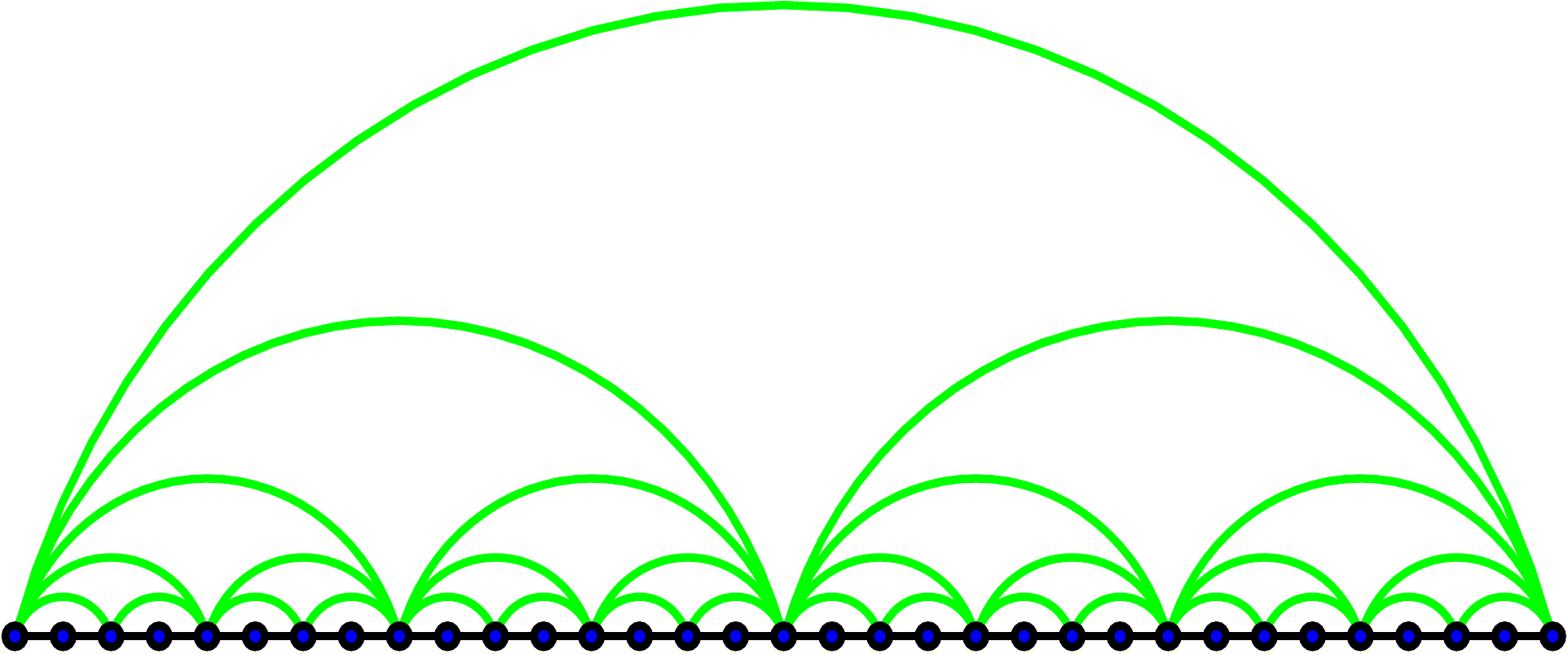}
\caption{\label{fig:Networks}Depiction of hyperbolic networks MK1 of generation
$k=5$. The recursive, hierarchical pattern is evident. The network
features a regular geometric structures, in form of a one-dimensional
backbone, and a distinct set of small-world links (shaded green arched lines).
Only the backbone bonds renormalize.}
\end{figure}

As a simple and generic member of the class of hyperbolic networks,
we consider here MK1, as depicted in Fig.~\ref{fig:Networks}. MK1
is recursively generated, starting with two sites connected by a single
edge at generation $k=0$. Each new generation combines two sub-networks
of the previous generation and adds single edge connecting the end
sites. As a result, the $k^{th}$ generation contains $N_k=2^{k}+1$ vertices,
$2^{k}$ backbone bonds, and $2^{k}-1$ small-world bonds. It is an
effectively one-dimensional version of the small-world Migdal-Kadanoff
hierarchical diamond lattice~\citep{Hinczewski06,Boettcher09c}, which
has been used previously to prove the existence of the discontinuous
transition in ordinary percolation~\citep{Boettcher11d}. Recently, this discontinuity has been studied for a variety of other hyperbolic networks for bond and site percolation in Ref.~\cite{Singh14}. 
Refs.~\citep{Nogawa12,PhysRevLett.108.255703} have used MK1 to study the $q$-state Potts
model for certain integer values of $q\geq2$. In principle, to relate the Potts model to percolation at $q\to1$ first requires a $q$-derivative of the partition function~\cite{Wu82}. However, we find that this derivative hardly affects the analytic properties, as we will show below by comparing each quantity obtained for $q\to1$ with the exactly known percolation results.   

In close correspondence to the Ising model on hyperbolic networks
discussed in Ref.~\citep{Boettcher10c} (see also Ref.~\citep{Nogawa12}),
we introduce the following couplings. All variables $x_{i}$ interact
with their nearest-neighbors along the backbone with a coupling $K_{0}$
(solid links in Fig.~\ref{fig:Networks}), while small-world neighbors
interact with a coupling $K_{1}$ (shaded links in Fig.~\ref{fig:Networks}).
Every $x_{i}$ also experiences a uniform external field $B$. Then,
in preparation for applying the renormalization group, the Potts-Hamiltonian
can be written as 
\begin{align}
-\beta\mathcal{H}=\sum_{n=1}^{N_k/2}\left(-\beta\mathcal{H}_{n}\right)+\mathcal{R}\left(K_{1}\right),
\end{align}
where $\mathcal{R}$ contains all remaining coupling terms of higher
level in the hierarchy while the $x_{i}$ on the backbone can be sectioned
into a sequence of three-variable graphlets. These consist of two
adjacent lattice-backbone bonds bridged by an arched small-world bond
in Fig.~\ref{fig:Networks}, each with their own ``sectional'' Hamiltonian,
\begin{align}
-\beta\mathcal{H}_{n}&=2I+K_{0}\left(\delta_{x_{n-1},x_{n}}+\delta_{x_{n},x_{n+1}}\right)+K_{1}\delta_{x_{n-1},x_{n+1}}\nonumber \\ &+B\left[\left(\delta_{1,x_{n-1}}+\delta_{1,x_{n}}\right)+\left(\delta_{1,x_{n}}+\delta_{1,x_{n+1}}\right)\right].
\end{align}
Here, $I$ is a constant that fixes the overall energy scale. We conveniently
choose new variables, similar to (inverse) ``activities\textquotedbl{}
\citep{Plischke94}, $C=e^{-2I}$, $\quad\kappa=e^{-K_{0}}$, and
$\theta=e^{-B}$. Similarly defined are the control parameters of temperature
$\mu=e^{-\beta J}$ and field $\eta=e^{-\beta h}$, with $\beta=1/kT$
and $h$ as the external field. The ``raw'' (unrenormalized) couplings
are considered as uniform: $K_{0}=K_{1}=\beta J$, where we fix the energy scale
via $J=1$, while $B=\beta h$. However, we find that $K_{1}$ does
not change under renormalization and retains its raw value, $e^{-K_{1}}=\mu$,
a key distinguishing feature of these hyperbolic networks~\citep{Boettcher10c,BoBr12}.
Thus, the initial values of the renormalizing activities are $C_{0}=1$,
$\kappa_{0}=\mu$, and $\theta_{0}=\eta$. Then, we can rewrite the
sectional Hamiltonian as
\begin{eqnarray*}
e^{-\beta\mathcal{H}_{n}} & = & C^{-1}\kappa^{-\left(\delta_{x_{n-1},x_{n}}+\delta_{x_{n},x_{n+1}}\right)}\mu^{-\delta_{x_{n-1},x_{n+1}}}\\
 &  & \quad\theta^{-\left[\left(\delta_{1,x_{n-1}}+\delta_{1,x_{n}}\right)+\left(\delta_{1,x_{n}}+\delta_{1,x_{n+1}}\right)\right]}.
\end{eqnarray*}
 The RG consists of successively tracing out the central variable
$x_{n}$ and expressing the renormalized activities $(C^{\prime},\kappa^{\prime},\theta^{\prime})$
in terms of their priors $(C,\kappa,\theta,\mu)$:
\[
\sum_{x_{n}=1}^{q}e^{-\beta\mathcal{H}_{n}}=\left(C'\right)^{-\frac{1}{2}}\left(\kappa^{\prime}\right)^{-\delta_{x_{n-1},x_{n+1}}}(\theta^{\prime})^{-\left(\delta_{1,x_{n-1}}+\delta_{1,x_{n+1}}\right)}
\]
Although the remaining variables $x_{n\pm1}$ present us with up to
$q^{2}$ relations between new and old quantities, only \emph{three}
relations are independent. We can obtain the RG-flow for
$(\theta^{\prime}, \kappa^{\prime})$ at size $N_{k+1}$ in terms of $(\theta, \kappa)$ at size~$N_k$,
\begin{subequations}
\begin{align}
\theta^{\prime} & =\theta\,\sqrt{\frac{\theta^{2}+\kappa^{2}+\left(q-2\right)\theta^{2}\kappa^{2}}{1+\left(q-1\right)\theta^{2}\kappa^{2}}},\\
\kappa^{\prime} & =\frac{\kappa\mu\left[1+\theta^{2}+\left(q-2\right)\theta^{2}\kappa\right]}{\sqrt{\left[1+\left(q-1\right)\theta^{2}\kappa^{2}\right]\left[\theta^{2}+\kappa^{2}+\left(q-2\right)\theta^{2}\kappa^{2}\right]}}.
\end{align}\label{eq:RGflow}
\end{subequations}
Note the \emph{explicit} dependence on the control parameter $\mu$
(temperature), characteristic for these networks~\citep{BoBr12}.

First, we analyze the RG-flow in Eq.~\eqref{eq:RGflow} in the absence
of an external field. For $\eta=1=\theta_{0}$, it remains $\theta_{k}=1$
for all $k>0$ and the RG-flow for the bond-coupling becomes 
\begin{align}
\kappa^{\prime} & =\kappa\mu\;\frac{2+(q-2)\kappa}{1+(q-1)\kappa^{2}}.\label{eq:RGflowNoH}
\end{align}
For the Ising model, $q=2$, Eq.~\eqref{eq:RGflowNoH} reduces with
$\kappa\hat{=}jk$ and $\mu\hat{=}j$ to the equivalent of Eq.~(6)
in Ref.~\citep{Nogawa12}, and for percolation, $q=1$, it reduces
with $\kappa\hat{=}1-T$ and $\mu\hat{=}1-p$ to Eq.~(1) in Ref.~\citep{Boettcher11d}. 

\begin{figure}
\includegraphics[bb=80bp 550bp 400bp 720bp,clip,width=1.3\columnwidth]{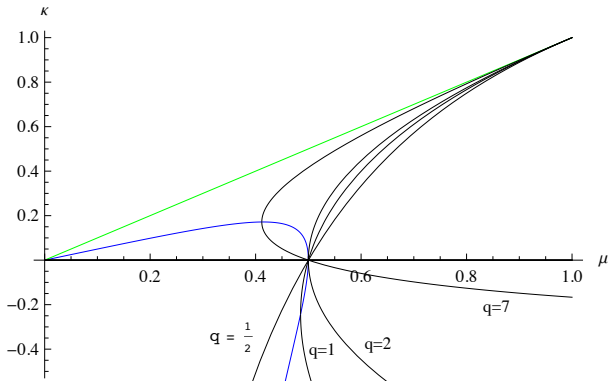}
\caption{\label{RG_soln} Plot of the fixed-point couplings $\kappa_{\infty}^{\pm}$
in Eq.~\eqref{eq:kappa_infty} as a function of temperature $\mu$
for values of $q=\frac{1}{2},1,2,7$ (black lines, from left to right
on the bottom). The blue-shaded line locates the branch-point singularity
in $\kappa_{\infty}^{\pm}$, evaluated at ${\cal D}_{q}\left(\mu\right)=0$
in Eq.~\eqref{eq:discriminant}, which starts from $\mu=0$ and $\kappa=-\infty$
at $q=1$ and rises to cross $\kappa=0$ exactly at $\mu=\frac{1}{2}$
for $q=2$, above which it recedes back to $\mu=0$ for $q\to\infty$.
The green-shaded line marks the raw coupling $\kappa_{0}=\mu$ from
which the RG-flow in Eq.~\eqref{eq:RGflowNoH} initiates and flows
vertically towards the nearest stable fixed-point, $\kappa_{\infty}^{+}>0$
or $\kappa_{\infty}^{0}=0$. For all $q>2$, the location of the branch
point provides the critical point $\mu_{c}\left(<\frac{1}{2}\right)$
with an infinite-order BKT transition~\citep{BoBr12}. For all $q<2$,
$\mu_{c}=\frac{1}{2}$ and the transition becomes discontinuous.}
\end{figure}

In the thermodynamic limit $(N_{k\to\infty})$, Eq.~\eqref{eq:RGflow} provides the fixed points $\kappa^{\prime}=\kappa=\kappa_{\infty}$:
\begin{eqnarray}
\kappa_{\infty}^{0}=0, & \qquad & \kappa_{\infty}^{\pm}=\frac{\mu}{2}\frac{\left(q-2\right)}{\left(q-1\right)}\left[1\pm\frac{\sqrt{{\cal D}_{q}\left(\mu\right)}}{\mu\left(q-2\right)}\right],\label{eq:kappa_infty}
\end{eqnarray}
where $\kappa_{\infty}^{+}>\kappa_{\infty}^{-}$,
with 
\begin{equation}
{\cal D}_{q}\left(\mu\right)=\mu^{2}\left(q-2\right)^{2}+4\left(2\mu-1\right)\left(q-1\right).\label{eq:discriminant}
\end{equation}
The strong-coupling limit $\kappa_{\infty}^{0}=0$ always exists for
any~$q$. However, the Ising case and the limit of Eq.~\eqref{eq:kappa_infty} that corresponds to percolation, both have to be considered with some care. Yet, we easily reproduce Eq.~(2) in Ref.~\citep{Boettcher11d}. We plot
the behavior of $\kappa_{\infty}^{\pm}$ in Fig.~\ref{RG_soln} as
a function of $\mu$ for a range of $q$ that highlights the peculiarities
of both cases. 

\begin{figure*}
\vskip -0.2in
\includegraphics[bb=80bp 640bp 210bp 740bp,clip,width=0.25\textwidth]{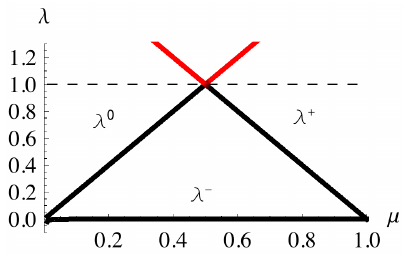}\includegraphics[bb=80bp 640bp 210bp 740bp,clip,width=0.25\textwidth]{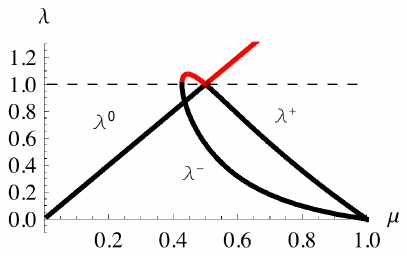}\includegraphics[bb=80bp 640bp 210bp 740bp,clip,width=0.25\textwidth]{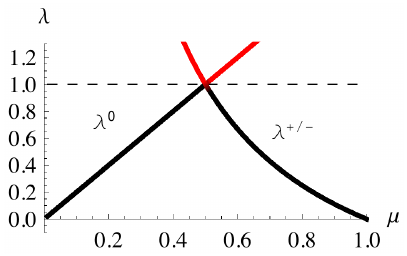}\includegraphics[bb=80bp 640bp 210bp 740bp,clip,width=0.25\textwidth]{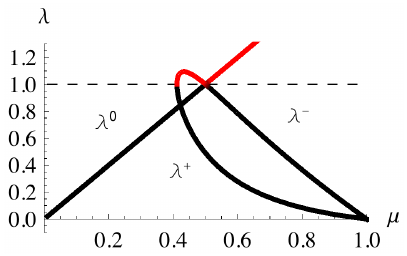}
\caption{\label{fig:lambda}Plot of the Jacobian eigenvalues $\lambda$ in
Eq.~\eqref{eq:lambdas} as a function of temperature $\mu$ for the
$q$-state Potts model on MK1 for $q=1,1.2,2,7$. For all $q$ at
the strong-coupling fixed point $\kappa_{\infty}^{0}=0$ we have $\lambda^{0}=2\mu$,
while the two branches $\lambda^{\pm}$ belonging to non-zero $\kappa_{\infty}^{\pm}$
change dramatically with $q$ and vary non-trivially with $\mu$.
At $q=1$, $\lambda^{-}=0$ and $\lambda^{+}=2-2\mu$ remain disconnected
but coalesce for $1<q<2$ (here, $q=1.2$) with a branch point at
unity (dashed line). At $q=2$, both branches degenerate into $\lambda^{\pm}=\left(1-\mu\right)/\mu$.
For $q>2$ (here, $q=7$), the branches re-open and resemble the case
$1<q<2$. Stable fixed points correspond to $\lambda<1$ (black),
unstable ones to $\lambda>1$ (red). }
\end{figure*}

To obtain the thermal exponent $y_{t}$, we calculate the relevant
eigenvalues along each branch of the fixed-point lines. The Jacobian
of the RG-flow in Eq.~\eqref{eq:RGflowNoH}, evaluated at $\kappa=\kappa_{\infty}$,
provides the eigenvalue 
\begin{equation}
\lambda_t=\left.\frac{\partial\kappa^{\prime}}{\partial\kappa}\right|_{\kappa_{\infty}}=\frac{2\mu\left[1+\left(q-2\right)\kappa_{\infty}-\left(q-1\right)\kappa_{\infty}^{2}\right]}{\left[1+\left(q-1\right)\kappa_{\infty}^{2}\right]^{2}}.\label{eq:Jacobian}
\end{equation}
Hence, with Eqs.~\eqref{eq:kappa_infty}, we get
\begin{eqnarray}
\lambda_t^{0} & = & 2\mu,\label{eq:lambdas}\\
\lambda_t^{\pm} & = & \frac{1-\mu}{q^{2}\mu}\left\{ 4\left(q-1\right)+\mu\left(q-2\right)^{2}\left[1\pm\frac{\sqrt{{\cal D}_{q}\left(\mu\right)}}{\mu\left(q-2\right)}\right]\right\} .\nonumber 
\end{eqnarray}
The behavior of the eigenvalues, depicted in Fig.~\ref{fig:lambda},
follows closely the description developed for general hyperbolic lattices
in Ref.~\citep{BoBr12}.

As Fig.~\ref{RG_soln} shows, for $q<2$ the critical point is located
always at $\mu_{c}=\frac{1}{2}$, such that the relevant thermal scaling
exponent~\citep{Pathria} $y_{t}=\log_{2}\lambda_t^{+}$ is given by
\begin{equation}
y_{t}\sim\frac{2\left(\mu_{c}-\mu\right)}{\ln2}\qquad\left(q<2,\mu\to\mu_{c}=\frac{1}{2}\right),\label{eq:yt_q<2}
\end{equation}
which indicates that $y_{t}$ vanishes for all $q<2$. As the generic
theory suggests~\citep{BoBr12}, this linear behavior is to be expected
when $\kappa_{\infty}^{+}$ is regular at $\mu=\mu_{c}$ such that
it possesses a linear slope, which is satisfied for $q<2$, see Fig.~\ref{RG_soln}. Eq.~\eqref{eq:Jacobian} implies that this will remain
true also for percolation, $q=1$, in particular. However, for the
Ising ferromagnet, and exactly \emph{only} at $q=2$, $\lambda_t$ in
Eq.~\eqref{eq:Jacobian}, and hence $y_{t}$, becomes a function of
$\kappa_{\infty}^{2}$. Miraculously, at $q=2$ the branch-point in
$\kappa_{\infty}$ is placed exactly on the $\kappa=0$-axis at $\mu_{c}$
with $\kappa_{\infty}^{\pm}\sim\pm\sqrt{2\left(\mu-\mu_{c}\right)}$.
Both effects combine into a linear correction but a different
coefficient, 
\begin{equation}
y_{t}\sim\frac{4\left(\mu_{c}-\mu\right)}{\ln2}\qquad\left(q=2,\mu\to\mu_{c}=\frac{1}{2}\right).\label{eq:yt_q=00003D2}
\end{equation}

\begin{figure*}
\vskip -0.4in
\includegraphics[bb=80bp 640bp 210bp 740bp,clip,width=0.25\textwidth]{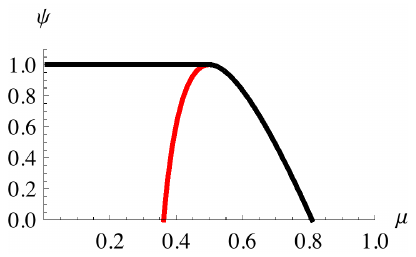}\includegraphics[bb=80bp 640bp 210bp 740bp,clip,width=0.25\textwidth]{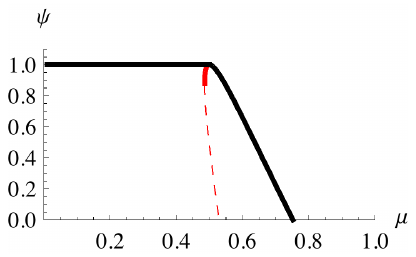}\includegraphics[bb=80bp 640bp 210bp 740bp,clip,width=0.25\textwidth]{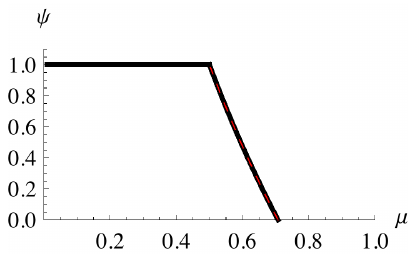}\includegraphics[bb=80bp 640bp 210bp 740bp,clip,width=0.25\textwidth]{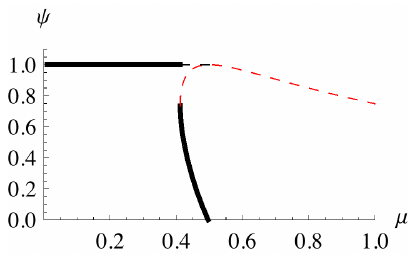}\caption{\label{fig:Psi_q}Plot of the exponent $\Psi$ as a function of temperature
$\mu$ for $q=1,\frac{3}{2},2$, and 7. The physical branch is given
by the thickened, dark lines. The red-shaded or dashed lines mark
the unphysical branches of $\Psi$. Note that $\Psi\left(\mu\right)$
is continuous at $\mu_{c}=\frac{1}{2}$ for $q<2$ with a parabolic
continuation but degenerates at $q=2$ (i.e., for the Ising model)
to a linear continuation. For $q>2$, $\Psi\left(\mu\right)$ is discontinuous
at $\mu_{c}=\mu_{B}<\frac{1}{2}$ and continues after the drop-off
on the lower branch of the root-singularity. } 
\end{figure*}

\begin{figure*}
\vskip 0in
\includegraphics[scale=0.6]{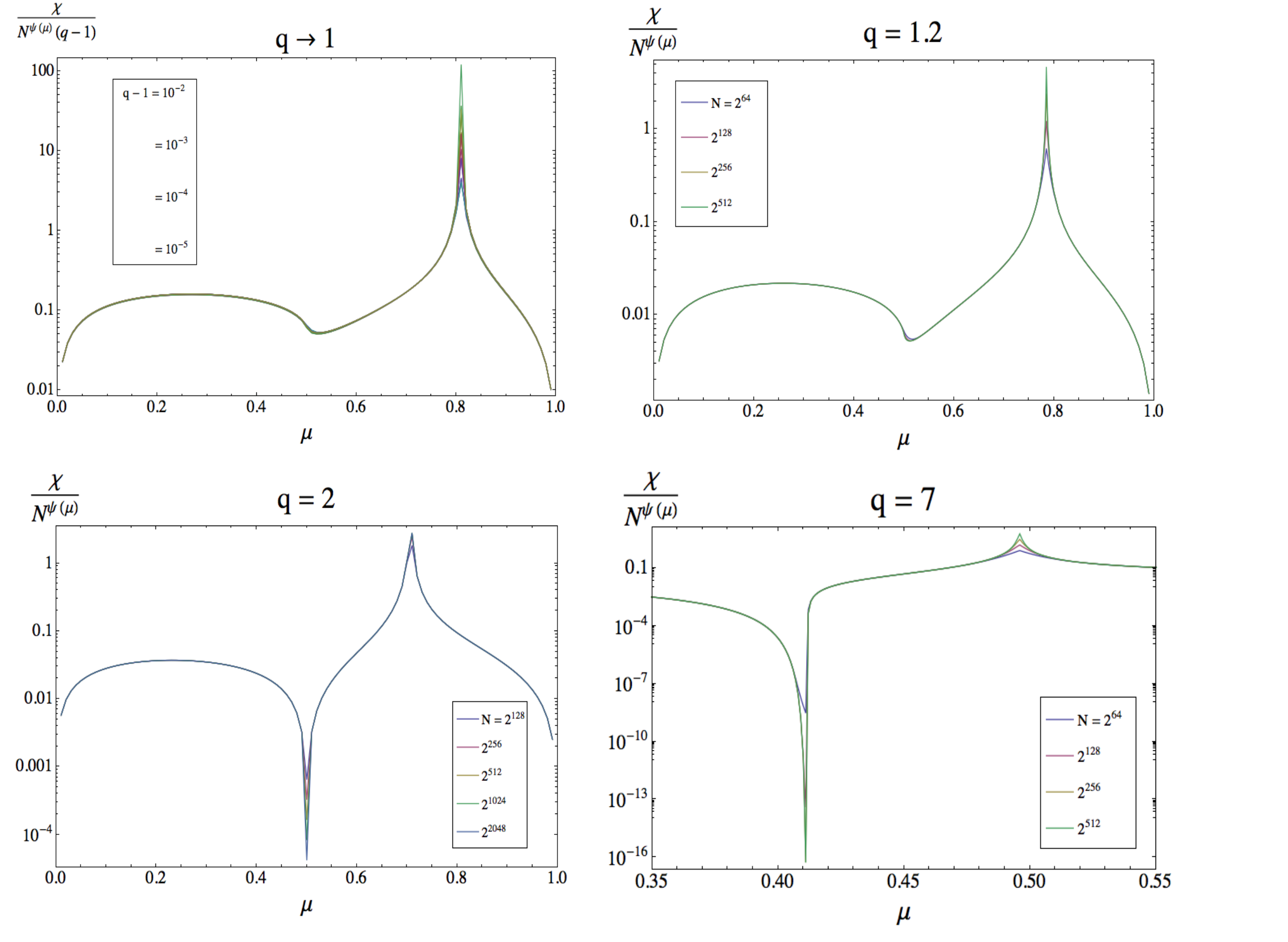}
\caption{\label{fig:chi_q}
Plot of the susceptibility $\chi$ as a function of temperature $\mu$, rescaled with the appropriate power of system size $N$ according to Eq.~(\ref{eq:chi}) using $\Psi(\mu)$ (and $\Psi\equiv1$ for $\mu<\mu_c$). From left to right, panels refer to $q\searrow1$, $q=1.2$, $q=2$, and $q=7$.
For the larger values of $q$, the rescaled $\chi$ are show for a sequence of increasing system sizes to demonstrate convergence in the thermodynamic limit, while for the  case $q\to1$ we show that $\chi$ requires a further rescaling by $1/(q-1)$ to converge to a finite value throughout (at some fixed, already large system size, here $N=2^{512}$). Note that in this scaling, the susceptibility develops a minimum at the corresponding value of $\mu_c$, however, it is sharply peaked at the  respective value at which $\Psi$ reaches (and then remains at) zero, see Fig.~\ref{fig:Psi_q}.}
\end{figure*}

For $q>2$, $\kappa_{\infty}^{+}$ no longer intercepts the strong-coupling
fixed-point line at $\kappa_{\infty}^{0}=0$, see Fig.~\ref{RG_soln},
rendering the issue moot. The critical behavior with an infinite-order
BKT-like transition is now described by analyzing the RG-flow near
the branch point itself,  as previously described~\citep{Hinczewski06,Boettcher10c,BoBr12,PhysRevLett.108.255703},
moving the critical point to $\mu_{c}=\mu_{B}<\frac{1}{2}$ for $q>2$,
where ${\cal D}_{q}\left(\mu_{B}\right)=0$ in Eq.~\eqref{eq:discriminant}
gives $\mu_{B}=2\left[q\sqrt{q-1}-2\left(q-1\right)\right]/\left(q-2\right)^{2}$. 

We now study the effect of an external field for $\eta\to1$ to analyze
the behavior of the order parameter near the transition. In the limit
$\eta\to1$, i.e., $\theta\to1$, and $\kappa\to\kappa_{\infty}$,
Eqs.~\eqref{eq:RGflow} provide the Jacobian matrix $\frac{\partial\left(\theta^{\prime},\kappa^{\prime}\right)}{\partial\left(\theta,\kappa\right)}$,
which turns out to be upper-triangular, since $\frac{\partial\theta^{\prime}}{\partial\kappa}=0$ at $\theta=1$.
Hence, the thermal exponent $y_{t}$ remains unaffected by the magnetic
field. Accordingly, we obtain the magnetic exponent $y_{h}=\log_{2}\lambda_{h}$
with
\begin{eqnarray}
\lambda_{h}&=&\frac{\partial\theta^{\prime}}{\partial\theta}=\frac{(q-2)\kappa_{\infty}^{2}+2}{(q-1)\kappa_{\infty}^{2}+1}\label{eq:lambda_h}\\
&=&\frac{2}{\mu q}+\frac{q-2}{2\mu q\left(q-1\right)} \left[\mu (3q-2)-\sqrt{{\cal D}_{q}\left(\mu\right)}\right],\nonumber
\end{eqnarray}
when evaluated along $\kappa_{\infty}^{+}$ near $\mu_{c}=\frac{1}{2}$.
As $\lambda_{h}$ is purely a function of $\kappa_{\infty}^{2}$,
there are no odd correction terms for the exponent near $\mu_{c}$,
\begin{equation}
y_{h}\sim1-\frac{8q}{\left(2-q\right)^{2}}\frac{\left(\mu_{c}-\mu\right)^{2}}{\ln2}\qquad\left(q<2,\mu\to\mu_{c}\right).\label{eq:yh_q<2}
\end{equation}
At $q=2$, we arrive at a more generic form, but merely by the fact
that $\kappa_{\infty}^{+}$ now has its root-singularity: 
\begin{eqnarray}
y_{h} & \sim & 1-2\frac{\left(\mu_{c}-\mu\right)}{\ln2}\qquad\left(q=2,\mu\to\mu_{c}\right).\label{eq:yh_q=00003D2}
\end{eqnarray}

Using the standard scaling relation~\citep{Pathria} for the behavior
of the order parameter near the transition,
\begin{equation}
m\sim\left(\mu_{c}-\mu\right)^{\beta},\quad\beta=\frac{1-y_{h}}{y_{t}},\label{eq:beta}
\end{equation}
we find $\beta=0$ for all $q<2$, while $\beta=\frac{1}{2}$ for
$q=2$. The later result, predicting a \emph{2nd}-order transition
with a mean-field-like exponent, has been obtained previously~\citep{Nogawa12}.
However, the former result suggests that the discontinuous percolation
transitions found on various hyperbolic networks~\citep{Boettcher11d,Nogawa13,Singh14}
is not an artifact of the particular percolation limit $q\to1$ of the
$q$-state Potts model but is maintained for all $q<2$. As soon as
$q>2$ (now with $\mu_{c}=\mu_{B}$), $1-y_{h}$ remains finite while
$y_{t}\to0$ for $\mu\to\mu_{c}$ such that $\beta$ diverges and
the transition becomes \emph{instantly} infinite-order, as explained
in Ref.~\citep{BoBr12}.

Another perspective on these peculiar transitions is provided by the
exponent~\citep{Nogawa12}
\begin{align}
\Psi=2y_{h}-1\label{eq:Psi}
\end{align}
for $\mu>\mu_{c}$, and $\Psi\equiv1$ for $\mu<\mu_{c}$, that describes
the divergence of the susceptibility with the system size,
\begin{equation}
\chi\sim N^{\Psi\left(\mu\right)},\label{eq:chi}
\end{equation}
not only near the transition but for all temperatures $0\leq\mu\leq1$.
For percolation, $\chi$ is related to the average size of the largest
cluster, $\left\langle S(p)\right\rangle $, after a derivative in $q$. In Fig.~\ref{fig:Psi_q},
we plot $\Psi\left(\mu\right)$ in Eq.~\eqref{eq:Psi} for various
$q$. The non-generic behavior of $y_{h}$ also manifests itself in
$\Psi$: no branch of $\Psi$ ever exceeds unity, as could be generically
expected, but rather $\Psi$ maintains a global extremum exactly of unit height
for all $q\not=2$. Also Ref.~\citep{Nogawa13} has found the same
``delicate'' behavior in $\Psi$ while studying the transformation
of the transition from discontinuous to infinite-order for a one-parameter
family of percolation models. The generic case can be observed, for
instance, for the families of Ising models studied in Ref.~\citep{Boettcher10c}
by applying an external field~\citep{BrBo14}, where $y_{h}$ and
hence $\Psi$ exhibit linear corrections near $\mu_{c}$ because the branch point singularity is outside the physical regime and $\Psi$ develops a maximum  at an unphysical value $\Psi>1$, rendering
the magnetization order parameter continuous at $\mu_c$.

While we found that the RG results for the couplings and $y_t$ of the $q$-state Potts are continuously connected to those of percolation at $q\to1$, this is not true for $\Psi$ (i.e., $y_h$ or possibly other quantities that are affected by a $q$-derivate of the partition function). However, a direct comparison reveals that the Potts model at $q=1$ in Eq.~(\ref{eq:yh_q<2}) and the percolation result from Eq.~(14) in Ref.~\citep{Boettcher11d} merely differ by a factor of 2 in their $2nd$-order correction near $\mu_c$. Their difference is more pronounced for larger  $\mu$, of course. For example, $\Psi$ remains positive for all  $\mu<1$ in percolation and only vanishes for zero bond-probability, $p=1-\mu=0$, suggesting that diverging clusters are always possible. In turn, $\Psi$ vanishes for non-trivial values of $\mu$ in the $q$-state Potts model, see Fig.~\ref{fig:Psi_q}, as is easily obtained from Eq.~(\ref{eq:lambda_h}). While insignificant for $q<2$, this remains true, remarkably, at $q\geq2$, which implies another potentially interesting transition in the behavior of $\chi$ at temperatures above criticality where fluctuations become independent of system size $N$, as shown in Fig.~\ref{fig:chi_q}. There, we demonstrate the collapse of the data for $\chi$ when properly rescaled with system size $N$ according to Eq.~(\ref{eq:chi}).  We observe a local minimum at $\mu_c$ which reaches a finite value on this scale for $q<2$ but appears to rapidly go to zero for all larger $q$ with increasing $N$. For $q\to1$, a further factor of $q-1$ is required to converge to a thermodynamic limit. (Remember that $\chi$ at $q=1$ is not directly related to the percolation cluster size, which would require an additional derivative of $q$ in the partition function.) Most remarkable is the spike in $\chi$ at the point $\mu$ where $\Psi(\mu)$ becomes zero, see Fig.~\ref{fig:Psi_q}.  Only above such temperature do fluctuations behave according to a high-temperature regime, independent of system size $N$. Such a transition from finite-size to diverging but subextensive fluctuations was absent, for instance, in the Ising model studied on any of the networks considered in Ref.~\citep{Boettcher10c}. 

To understand what makes these networks behave in such a non-generic
way, let us consider the case of percolation on a binary tree, the
most extreme hyperbolic network. Starting from the root of the tree
at level $i=0$, the average number of sites at branching level $i>0$
that are connected in a single cluster with the root is $\left\langle n_{i+1}\right\rangle =2p\left\langle n_{i}\right\rangle $,
hence, the average size of the rooted cluster at system size $N_{k}=2^{k+1}-1$
is $\left\langle S_{k}(p)\right\rangle =\sum_{i=0}^{k}\left\langle n_{i}\right\rangle =\left[(2p)^{k+1}-1\right]/\left(2p-1\right)$,
or $\left\langle S(p)\right\rangle \sim N^{\Psi\left(p\right)}$ with
$\Psi\left(p\right)=\log_{2}\left(2p\right)$ for $\frac{1}{2}<p<p_{c}=1$.
Just like on any other hyperbolic network, there are diverging, yet
sub-extensive, clusters well below $p_{c}$. (In fact, such diverging
cluster emerge as soon as there are spanning clusters connecting the
root to the perimeter of the tree, for $\frac{1}{2}<p<p_{c}$.) However,
here we find linear corrections,
\begin{equation}
\Psi\left(p\right)\sim1-\frac{1}{\ln2}\left(p_{c}-p\right)\qquad\left(p\to p_{c}\right),\label{eq:TreePsi}
\end{equation}
remembering that $\mu\hat{=}1-p$. We suggest that the relevant feature
that distinguishes the networks with non-trivial explosive percolation
transitions~\citep{Cho14} with $p_{c}<1$ from the trivial cases
with $p_{c}=1$, like this tree, is the interplay between a geometric
backbone and the small-world hyperbolic links. For instance, adding
lateral links turn the tree into a non-amenable graph with a non-trivial
transition, studied in Refs.~\citep{Nogawa09,Hasegawa10b}. Similarly,
MK1 here, or the hyperbolic Hanoi networks in Ref.~\citep{Singh14},
become non-trivial by mixing geometric with hyperbolic structures.
It is remarkable, that in Ref.~\citep{Hasegawa13c} the tree-approximation
for $\Psi\left(p\right)$ for percolation on a hyperbolic network,
that is exact for $p<p_{l}$ but predicts a trivial transition at
$p_{c}=1$ with linear corrections in $\Psi\left(p\right)$, breaks
down above $p_{l}$ exactly where the geometric backbone begins to
facilitate spanning clusters~\citep{Boettcher09c} and ultimately
leads to a non-trivial explosive transition at $p_{c}<1$, again with
quadratic corrections in $\Psi\left(p\right)$~\citep{Singh14}. While
renormalization group calculations provide these exact results, it
will likely require more detailed numerical simulations or better
models to elucidate the role of geometry in the dynamics of cluster
formation that lead to such corrections.

Finally, we speculate on the origin of the singularity of the $q$-state
Potts model for $q\to2$. What distinguishes the $q\geq2$ case from
that for $q<2$ is the nature of the $Z_{q}$-symmetry~\citep{Wu82}.
While for $q<2$ (and, correspondingly, for percolation) it is sufficient to merge
two clusters by merely adding a link, for $q\geq2$ (i.e., for the
Ising spin model or other multi-state degrees of freedom) two cluster
only merge when both share the same orientation. Apparently, the analytic
continuation in $q$ delimits these two distinct cluster-growth processes
via a singularity at $q\to2$.

We like to thank T. Nogawa and T. Hasegawa for fruitful discussions.
This work was supported by NSF (DMR \#1207431 \& IOS \#1208126) and James S. McDonnell Foundation (\#220020321).

\bibliography{Boettcher}
 
\end{document}